\documentclass[prl,twocolumn,showpacs,superscriptaddress]{revtex4-1}

\usepackage{graphicx}
\usepackage{setspace}
\usepackage{amsmath}
\usepackage{mathtools}
\usepackage{tabularx}
\usepackage{mathrsfs}
\usepackage{xcolor}
\newcommand{\be}{\begin{equation}}
\newcommand{\ee}{\end{equation}}
\newcommand{\bea}{\begin{eqnarray}}
\newcommand{\eea}{\end{eqnarray}}

\begin{document}

\title{Superfluid-Insulator Transition unambiguously detected by entanglement in one-dimensional  disordered superfluids}

\author{G. A. Canella}
\author{V. V. Fran\c{c}a}
\email{vivian.franca@unesp.br}
\affiliation{Institute of Chemistry, S\~{a}o Paulo State University, 14800-090, Araraquara, S\~{a}o Paulo, Brazil}

\begin{abstract}

We use entanglement to track the superfluid-insulator transition (SIT) in disordered fermionic superfluids described by the one-dimensional Hubbard model. Entanglement is found to have remarkable signatures of the SIT driven by {\it i)} the disorder strength $V$, {\it ii)} the concentration of impurities $C$ and {\it iii)} the particle density $n$. Our results reveal the absence of a critical potential intensity on the SIT driven by $V$, i.e. any small $V$ suffices to decrease considerably the degree of entanglement: it drops $\sim 50\%$ for $V=-0.25t$. We also find that entanglement is non-monotonic with the concentration $C$, approaching to zero for a certain critical value $C_C$. This critical concentration is found to be related to a special type of localization, here named as {\it fully-localized state}, which can be also reached for a particular density $n_C$. Our results show that the SIT driven by $n$ or $C$ has distinct nature whether it leads to the full localization or to the ordinary one: it is a first-order quantum phase transition when leading to full localization, and a smoother transition when reaching ordinary localization. In contrast, the SIT driven by $V$ is always a smoother transition independently on the type of localization reached.\end{abstract}


\maketitle

Localization and superconductivity are phenomena with opposite features. In a superconductor the effective attractive interactions lead to 
long-range electronic order, allowing supercurrents without resistance. In contrast, localized systems have no flux of electrical charge, 
thus behaving as insulators. Hence nanostructures, ultracold atoms and condensed-matter systems exhibiting both, superconductivity and 
localization, represent a rich scenario with interesting and unconventional properties. 

According to Anderson's theorem \cite{anderson}, localization is expected to occur in strong or moderate disordered 
systems. In the regime of strong disorder, the electronic wavefunctions become localized, transforming metals or superconductors in 
insulators \cite{supLoc}. The latter case $-$ the so-called superfluid to insulator transition (SIT) 
\cite{chinesRef12}  $-$ should be driven either by increasing the disorder strength or by 
lowering the particle density. The SIT is marked 
by the decrease of the superfluidity fraction \cite{benjaminRef1} and a distinct non-monotonic behavior of the 
condensate fraction with disorder intensity \cite{crossoverBCSBEC}.

Disorder remains a current topic, with possible implications to quantum information technologies 
\cite{mbl2019}, while SIT driven by disorder is very important for being considered a paradigmatic quantum phase transition in many-body systems 
\cite{benjamin}. Understanding the SIT can reflect in better comprehension of many complex 
superfluid systems such as high-$T_c$ superconductors (intrinsically disordered \cite{chinesRef12}), superconducting nanowires, amorphous superconductors, ultracold atomic gases, and molecules 
\cite{benjaminR}. 

Most of the experiments 
have focused on weak or non-interacting bosonic systems
\cite{ref2-32}. Theoretically, one faces the challenge of treating many-body 
interactions and the many realizations typically needed for describing disorder.  Although the properties of the SIT should not depend on the entity used to track the transition \cite{carterDisorder}, some 
quantities may not be sufficiently sensitive to the SIT, and thus do not present any distinct behavior.

In this sense entanglement 
measures appear as good candidates to pursue the SIT, since they have been successfully used to detect quantum phase transitions in 
several contexts \cite{qpt}. Localization has been investigated 
via 
entanglement in metals \cite{berkovitsRef13}, bosonic systems 
\cite{islamRef18}, Bose-Fermi mixtures \cite{albusRef37} and spinless fermions 
\cite{royRef14}. 

However, none of the previous studies {\it i)} has considered purely fermionic superfluid systems and {\it ii)} none has found entanglement as an unambiguous 
signature of the SIT. Furthermore, important features of the SIT are in current debate, as the existence or not of a critical disorder intensity for localization 
in 1D and 2D systems \cite{chinesRef12, buchholdRef16}, as well as 
the nature of the transition, whether it is a more pronounced transition or a crossover \cite{benjaminRef1,bouRef4}. 

In this work we report entanglement as an unequivocal signature of the SIT in one-dimensional disordered fermionic superfluids. We 
explore a unique opportunity to better understand the SIT through standard density-functional theory (DFT) \cite{dft} applied to the Hubbard model,
within a specially designed density functional for the linear entropy \cite{francaAmico2011}. This approach allows us to generate many realizations for each set of parameters, comprising a huge amount of data, what would be prohibitive via exact methods as density-matrix renormalization group. 

Our main results can be summarized 
as follows: (a) Entanglement detects the SIT driven by all the parameters: disorder strength,  concentration of impurities and particle density. (b) The SIT driven by the disorder strength requires no critical potential intensity to occur: any small disorder suffices to drive the transition. (c) For sufficiently strong disorder strength, a particular type of localization emerges at a critical $C_C$ or at a critical $n_C$. This state is marked by null entanglement and therefore is here named as {\it fully-localized state}. (d) The SIT driven by the concentration and by the particle density are found to be {\it i)} quantum phase transitions of first order whenever the system reaches the fully-localized state and {\it ii)} smoother transition, when the system reaches ordinary localization. In contrast, the SIT driven by the disorder strength is smoother for both, ordinary and full localization.

\begin{figure}[t]
 \centering
  \includegraphics[scale=0.29]{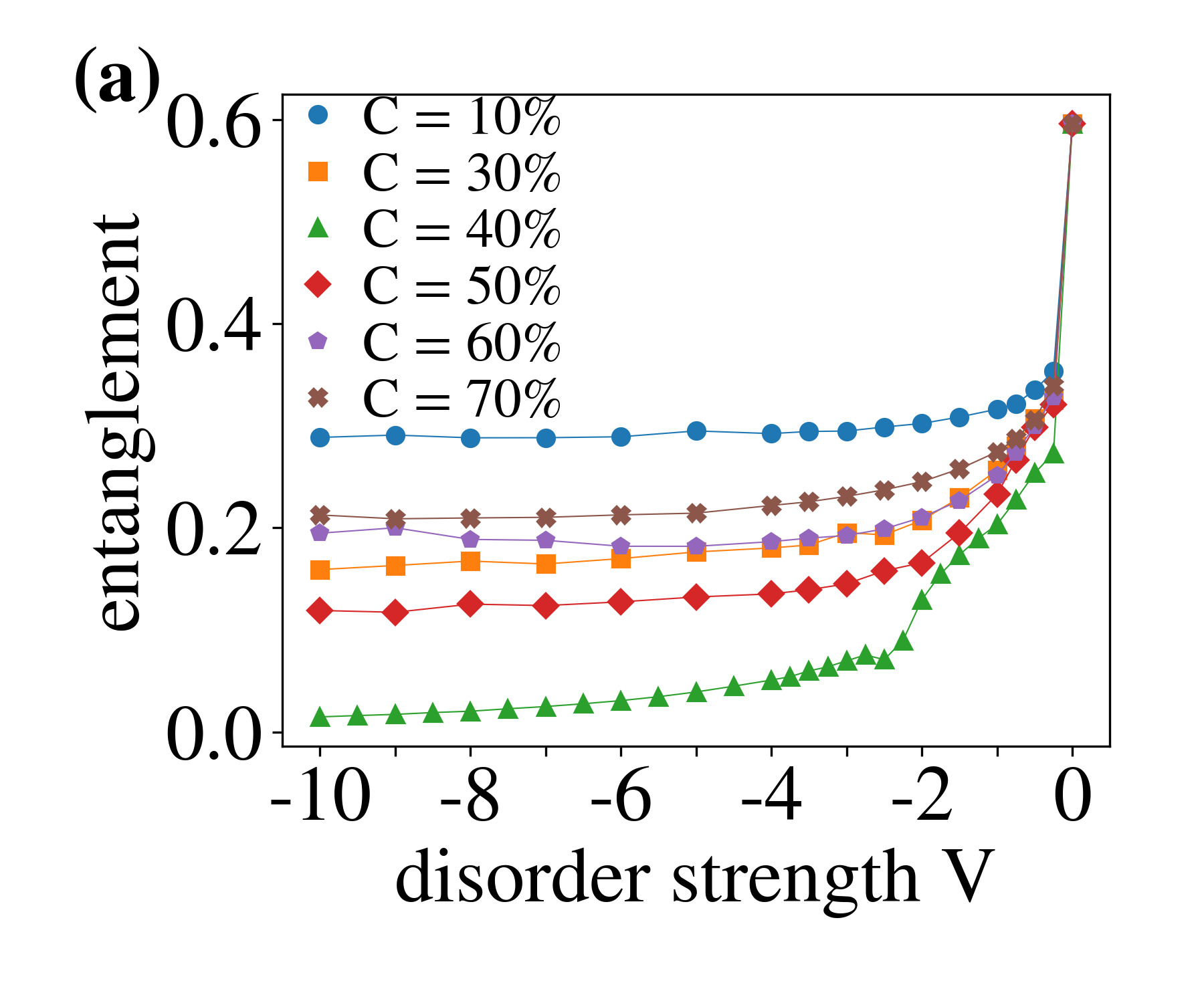} \hspace{-0.4cm}\includegraphics[scale=0.29]{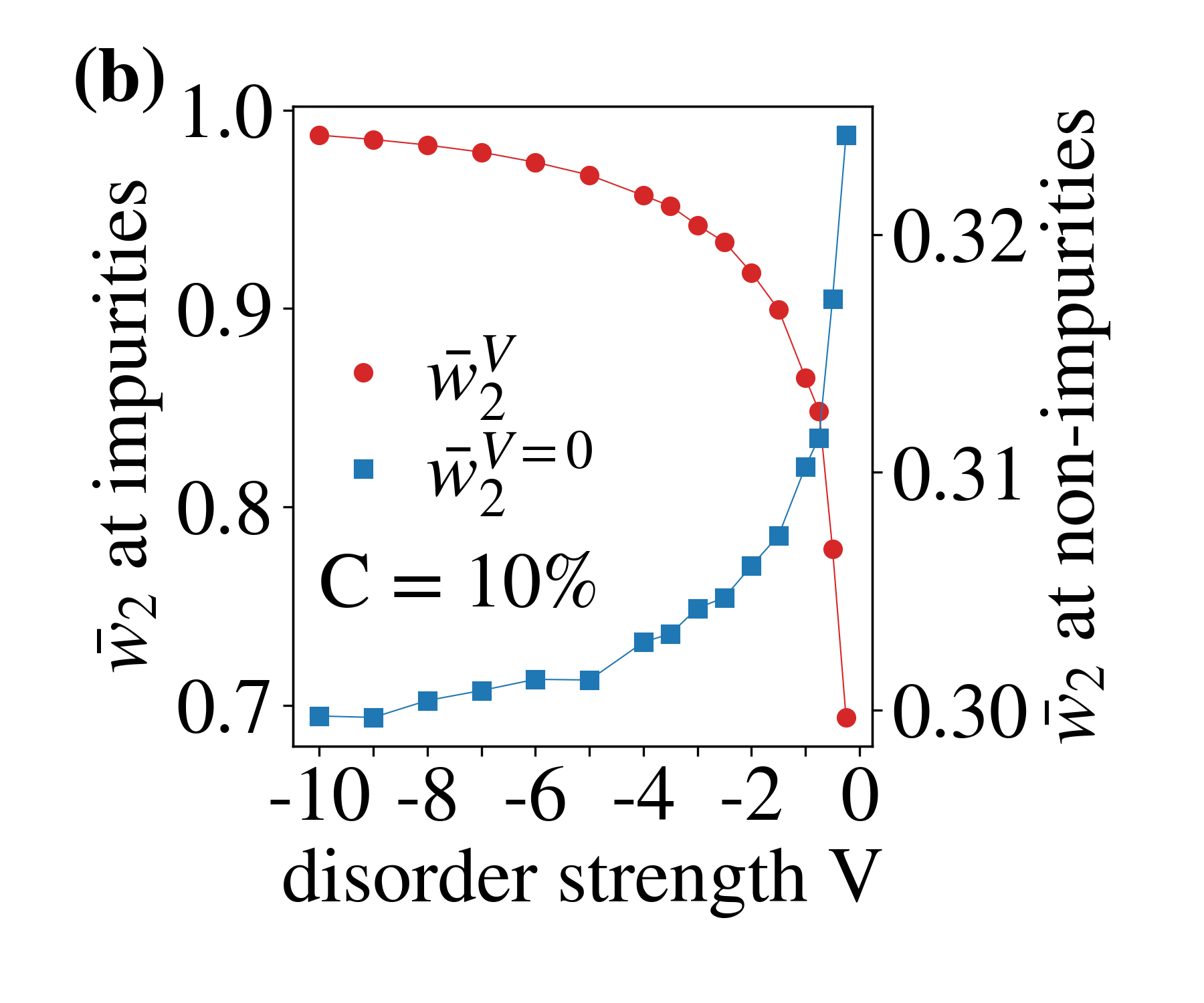}\\
       \vspace{-0.2cm}
    \includegraphics[scale=0.29]{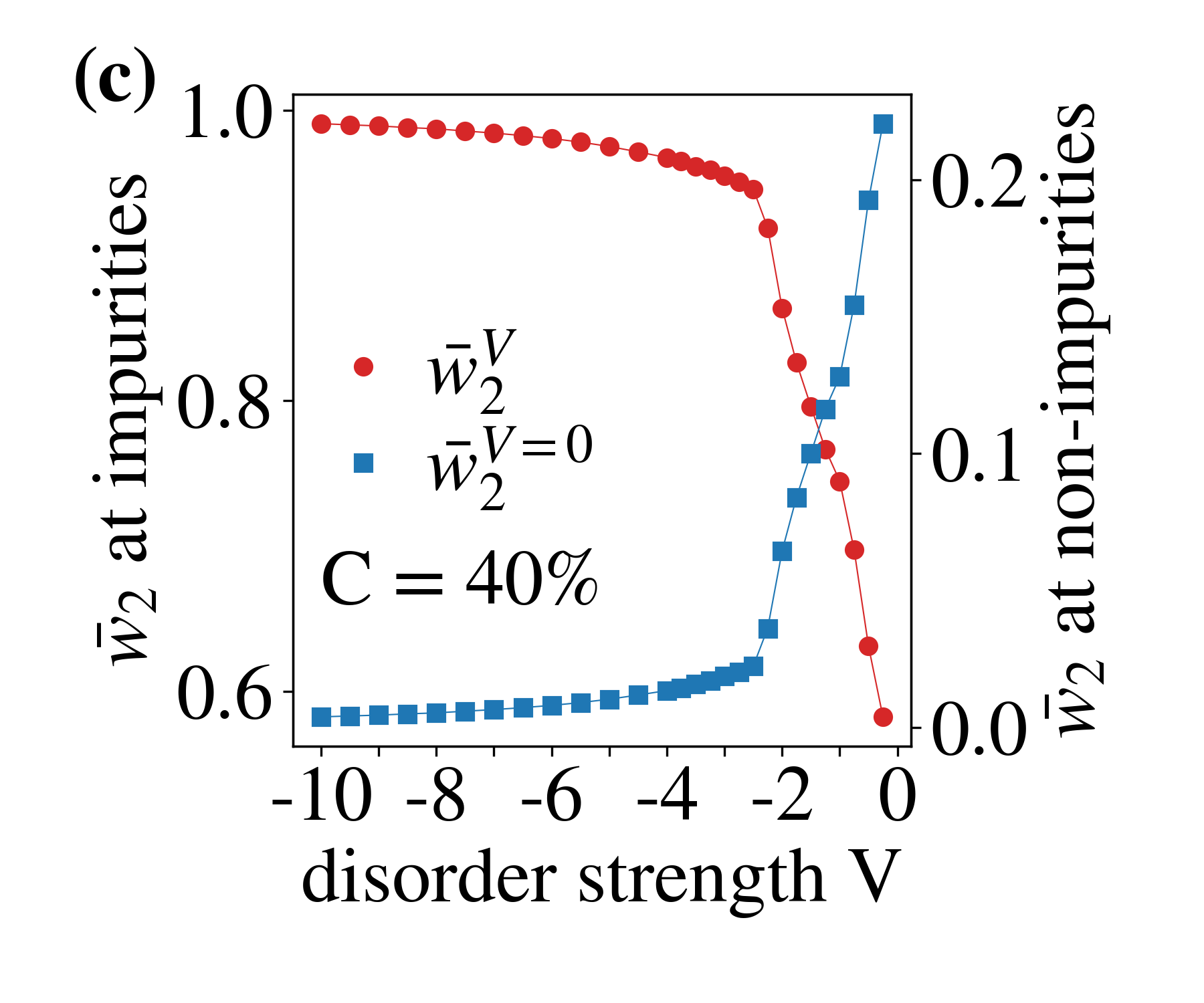}\hspace{-0.4cm} \includegraphics[scale=0.29]{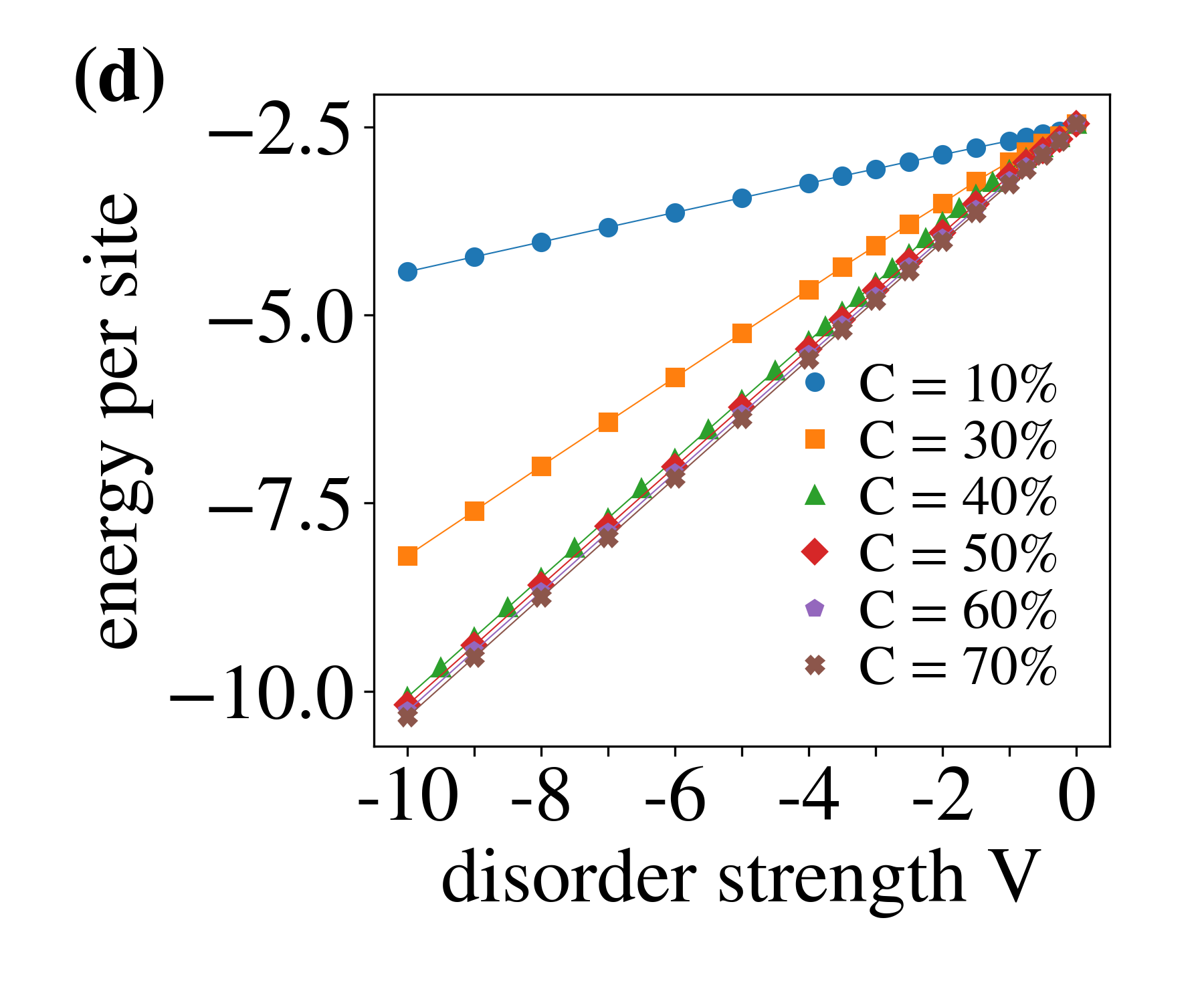}\\
    \vspace{-0.4cm} 
           \caption{SIT driven by the disorder strength $V$: entanglement $\mathcal{L}$ (a), average double-occupation probabilities at impurity ($\bar w_2^{V}$) and non-impurity ($\bar w_2^{V=0}$) sites for $C=10\%$ (b) and $C=40\%$ (c), and per-site ground-state energy (d).}\label{fig1}
   \end{figure}

We consider one-dimensional superfluids at zero temperature, where classical fluctuations are absent, described by the fermionic 
Hubbard model:

\begin{eqnarray}
H=-t\sum_{<ij>\sigma}\hat c_{i\sigma}^{\dagger}\hat c_{j\sigma}
 +U\sum_{i}\hat{n}_{i\uparrow}\hat{n}_{i\downarrow}+\sum_{i\sigma}V_i\hspace{0,1cm}\hat{n}_{i\sigma}.
\label{eqn:HubbardHamiltonian}
\end{eqnarray}
Here $U<0$ is the attractive on-site interaction, $t$ is the hopping parameter between neighbor sites $<ij>$ and $V_i$ is the external potential, used to 
simulate the disorder.  The density operator is $\hat n_{i,\sigma}=\hat c_{i,\sigma}^{\dagger}\hat c_{i,\sigma}$, where $\hat c_{i,\sigma}^{\dagger}$ ($\hat c_{i,\sigma}$) is the creation (annihilation) operator of fermionic particles with 
$z$-spin component $\sigma=\uparrow,\downarrow$ at site $i$,  $n=N/L$ is the average density or 
filling factor, $N=N_\uparrow+N_\downarrow$ the total number of particles and $L$ the chain size. We consider spin-balanced populations, 
$N_\uparrow=N_\downarrow$, $L=100$, open boundary conditions, $U=-5t$ and $n=0.8$, unless otherwise stated.

Our disordered superfluids are characterized by a certain concentration $C$ of pointlike impurities 
\cite{buchholdRef16,francaAmico2011} of strength $V$ randomly placed along the chain. For each 
set of parameters, $(C, V; U, n)$, we generate $M=100$ samples to avoid features due to specific impurity configurations. Then any property for that given set of parameters is averaged over
$M$ \cite{sup1}. This implies however in 
a huge amount of data, which would be prohibitive with numerically exact methods, such as density-matrix renormalization group calculations. 

We apply instead standard DFT techniques to solve the model 
$-$ the self-consistent Kohn-Sham scheme within a local density approximation for the exchange-correlation energy  \cite{dft} $-$ obtaining the ground-state energy and the density profile, with 
fair precision ($\sim 1\%$ for chains of this size). To quantify the degree of entanglement in such superfluids we adopt a specially designed density functional for the linear entropy $\mathcal{L}$ 
\cite{francaAmico2011}, which is then used as input in a local 
density approximation, following the protocol proposed in Ref.\cite{prl2008}, to obtain the average single-site entanglement of the inhomogeneous disordered chains. 
   
The SIT is expected to be driven either by enhancing the disorder strength (at a constant density) or by lowering the particle density (at a constant 
disorder) \cite{chinesRef12}. Nevertheless, in our pointlike type of disorder 
\cite{buchholdRef16,francaAmico2011} (see Refs. \cite{islamRef18, albusRef37, royRef14, praRef25}
for other potential landscapes) the concentration of impurities is an additional important parameter whose impact on the transition remains to be investigated. Thus we 
analyze the SIT tuned by disorder strength, impurities' concentration and particle density. 

\begin{figure}
 \centering
      \includegraphics[scale=0.29]{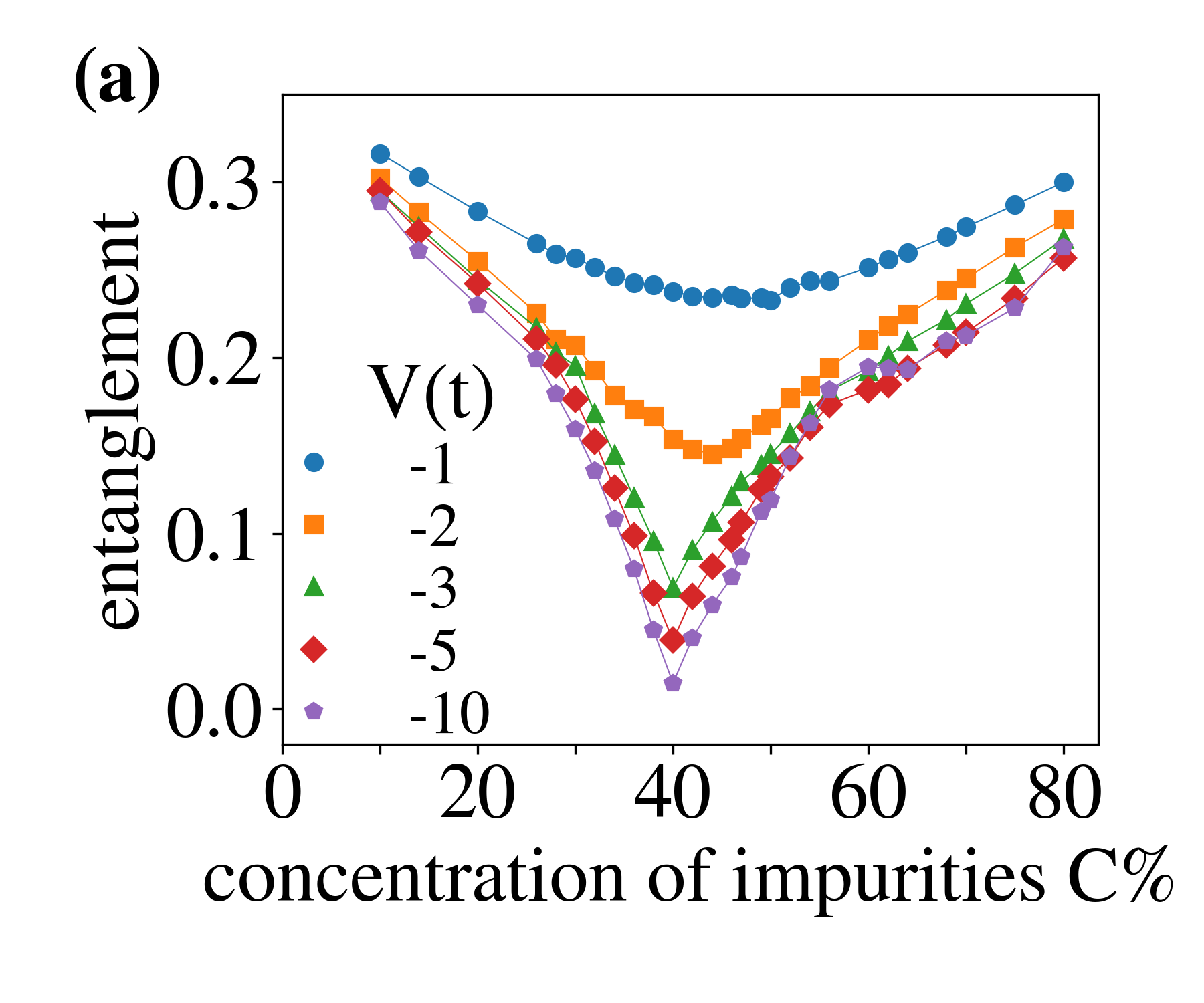} \hspace{-0.4cm}\includegraphics[scale=0.29]{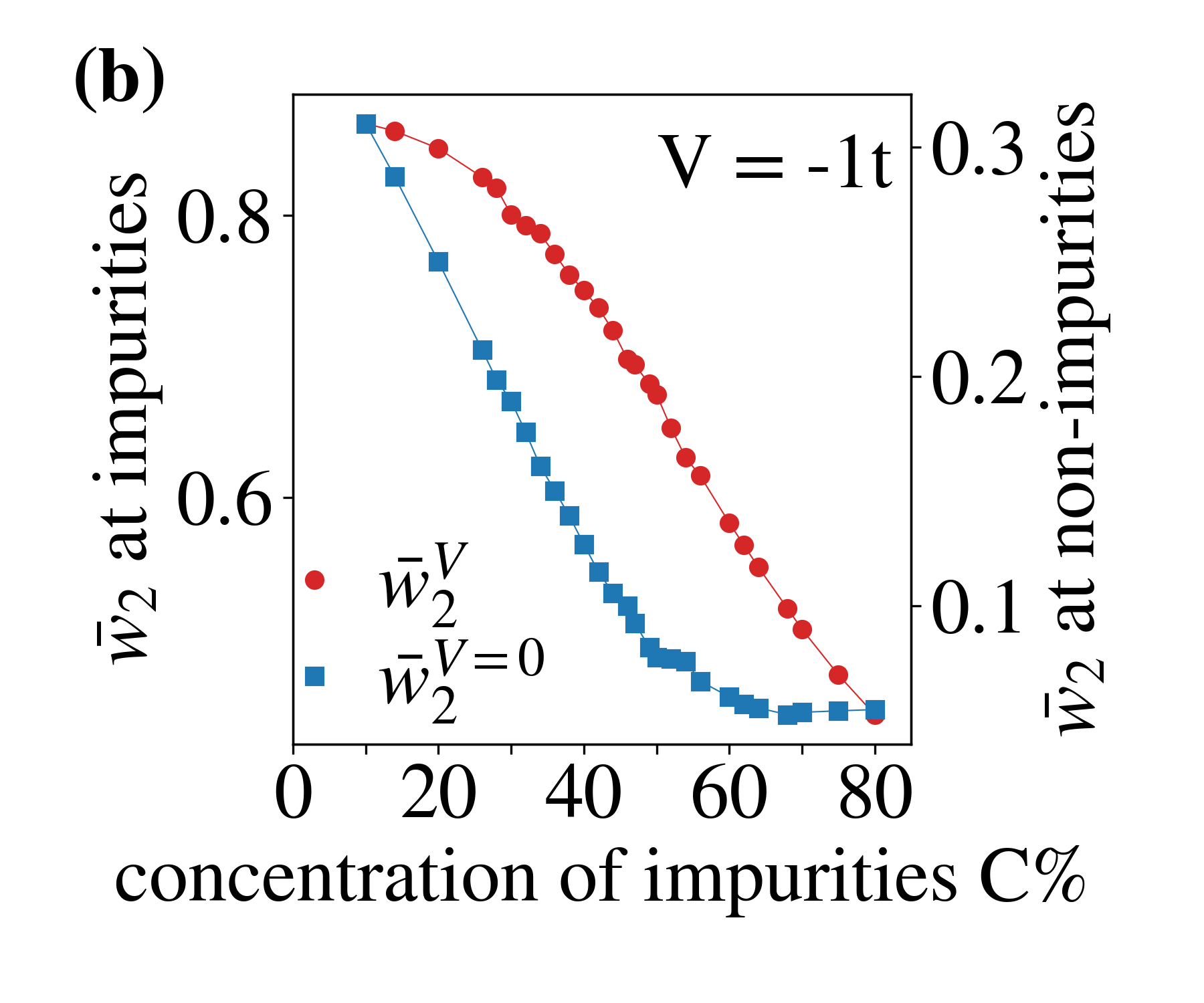}\\
       \vspace{-0.2cm}
    \includegraphics[scale=0.29]{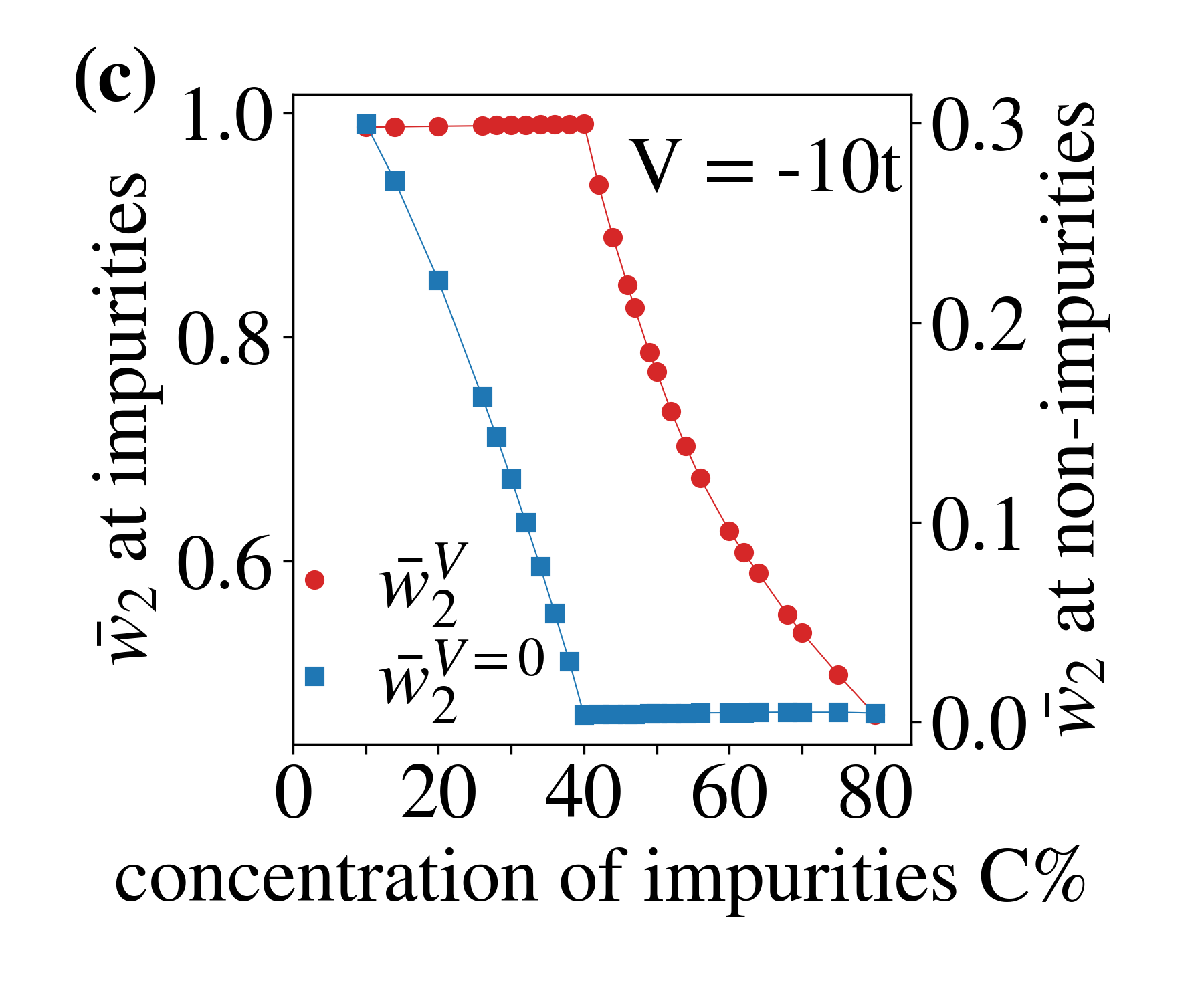}\hspace{-0.4cm} \includegraphics[scale=0.29]{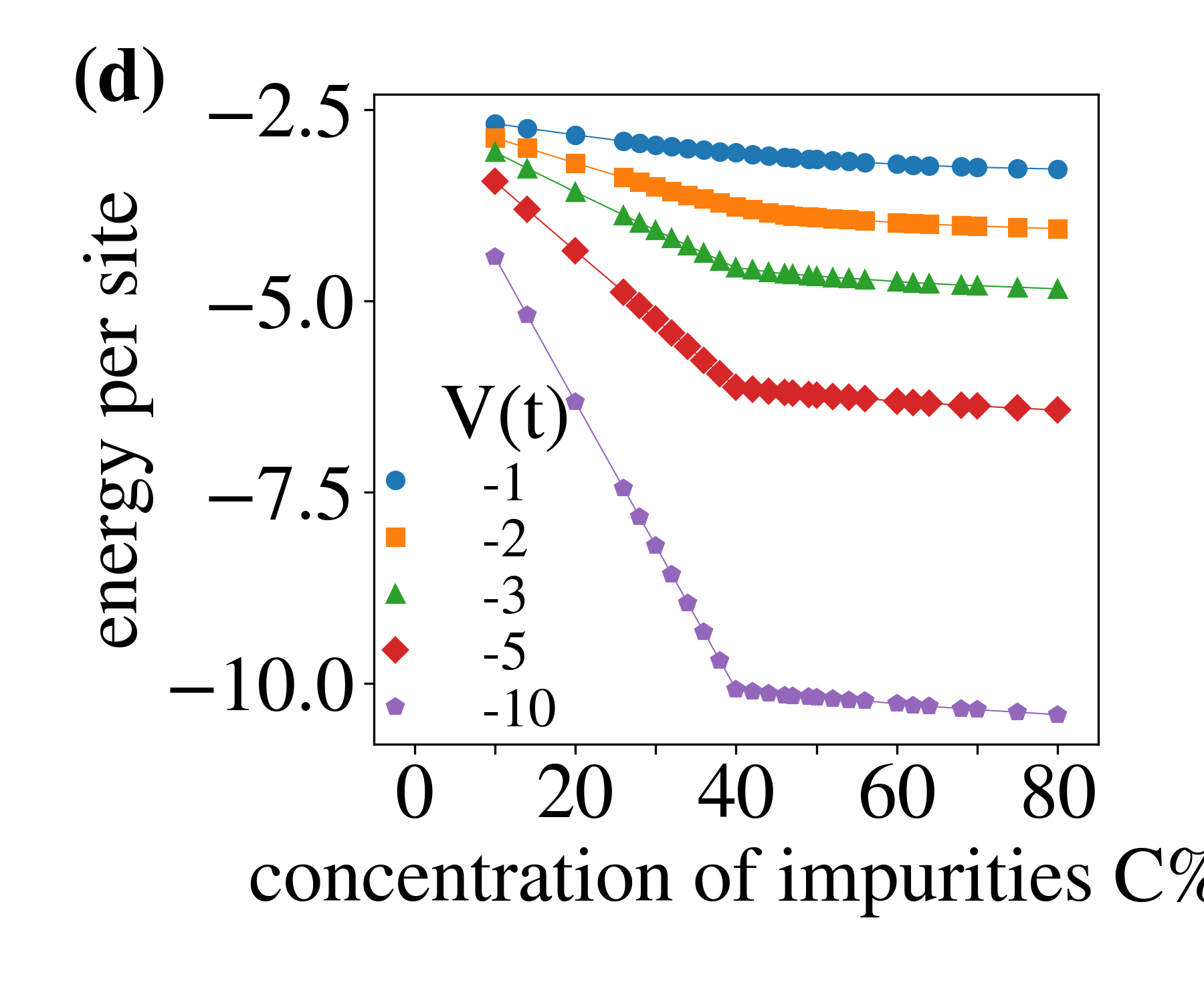}\\
    \vspace{-0.4cm}
      \caption{SIT driven by the impurities' concentration $C$: entanglement $\mathcal{L}$ (a),  average double-occupation probabilities at impurity ($\bar w_2^{V}$) and non-impurity ($\bar w_2^{V=0}$) sites for $V=-t$ (b) and $V=-10t$ (c), and per-site ground-state energy (d).}\label{fig2}
   \end{figure}

{\it SIT driven by disorder strength.}$-$We start by tracking the transition across the impurities' strength $V$ at a fixed density for several concentrations.  Figure \ref{fig1}-a reveals that entanglement quickly saturates with $V$: $\mathcal{L}$ drops already $\sim 50\%$ for $V= -0.25t$ for any $C$. This entanglement saturation is an unequivocal signature of the SIT: the initial superfluid, at $V=0$, is transformed into strongly-coupled localized dimers, which now make the system an insulator with no further effect by enhancing $|V|$. Also, the quick decreasing of $\mathcal{L}$ for small $V$ supports the absence of a critical disorder intensity for the SIT driven by disorder strength: even a weak potential drives the system from the superfluid state to the localized one.

We also find that entanglement is non-monotonic with $C$: from its maximum at $C=0$ (equivalent to $V=0$), entanglement decreases with $C$ reaching its minimum at a certain critical value $C=C_C=40\%$, and thus increasing for $C>C_C$. Although any concentration leads the system to localization, two features suggest that at $C=C_C$ there is a special type of localization: $\mathcal{L}$ behaves differently from all the other $C's$ for weak disorder ($-3t<V<0$) and it is the only case where $\mathcal{L}\rightarrow 0$ for $V\rightarrow-\infty$. Indeed we find that $C_C$ corresponds exactly to half of the total number of particles at the system, $C_C=100(n/2)\%$ (this relation is valid in general, see \cite{pra}), indicating that the strongly coupled dimers are {\it fully localized} at the impurities sites, which are the most attractive ones.

This is confirmed by the average double-occupation probabilities at impurity sites ($\bar{w}_2^{V}=\sum_i\langle \hat{n}_{i\uparrow}\hat{n}_{i\downarrow}\rangle/L_V $) and non-impurity sites ($\bar{w}_2^{V=0}=\sum_i\langle \hat{n}_{i\uparrow}\hat{n}_{i\downarrow}\rangle/(L-L_V) $): For $C\leq C_C$ (Fig.\ref{fig1}-b), the dimers are in larger number than impurity sites, thus there are localized dimers at impurity sites ($\bar{w}_2^{V}\rightarrow 1$  for $V\rightarrow -\infty$) but also some delocalized pairs at non-impurity sites ($\bar w_2^{V=0}$ saturates in a finite value, $\bar w_2^{V=0}(V\rightarrow -\infty)\approx 0.3$). For $C\geq C_C$ (Fig.\ref{fig1}-c), the system remains with a certain degree of delocalization since there are more impurity sites than dimers, but the non-impurity sites are empty ($\bar w_2^{V=0}\rightarrow 0$ for $V\rightarrow -\infty$). 

Thus the full localization at $C_C$ represents a well defined state with unitary probability of pairs at impurity sites and null probability of pairs at non-impurity sites, which is then characterized by $\mathcal{L}\rightarrow 0$, in contrast to the ordinary localization in which $\mathcal{L}$ saturates, but at finite values.  Interestingly though the ground-state energy, shown in Fig.\ref{fig1}-d, has no special behavior \cite{sup}, thus suggesting that the SIT driven by $V$ is smooth, a second-order quantum phase transition or simply a crossover \cite{sachdev}, independently on which type of localization is reached.

{\it SIT driven by concentration of impurities.}$-$Next we monitor entanglement across the disorder concentration, as shows Figure \ref{fig2}-a for weak, moderate and strong disorder 
intensities. We find that a minimum disorder strength $V_{min}$ is needed for the existence of the full localization at the critical $C_C$, marked by a sudden non-monotonicity of the entanglement and by $\mathcal{L}\rightarrow 0$  for $V\rightarrow -\infty$. For 
$|V|<V_{min}$ (in this case $V_{min}\sim3t$), we observe a very distinct behavior of entanglement as a function of $C$: the non-monotonicity 
is not abrupt, the minimum of entanglement is larger than zero and it does not occur at $C=C_C=40\%$. 

We interpret this distinct behavior for $|V|<V_{min}$ as a {\it frustrated full localization}:  $V$ is not sufficiently strong 
to fully localize the dimers at the impurity sites. This is indeed confirmed by the average double occupancy at impurity and non-impurity sites, shown in Fig.\ref{fig2}-b: $\bar w_2^{V=0}$ is always above zero for $V=-t$. In contrast, Fig.\ref{fig2}-c shows that for $V=-10t$ the fully-localized state is reached at $C_C$: $\bar w_2^{V}\approx 1$  for $C\leq C_C$, and $\bar w_2^{V=0}\rightarrow 0$ for $C\geq C_C$. 

For $|V|>V_{min}$, the ground-state energy, shown in Figure \ref{fig2}-d, reveals the nature of the transition from the ordinary to the fully-localized state driven by $C$: it has a discontinuous first derivative at $C=C_C$ \cite{sup}, thus 
characterizing a first-order quantum phase transition \cite{sachdev}. For $|V|<V_{min}$ the energy is smooth, with no discontinuity 
neither on the first nor on the second derivative \cite{sup}, suggesting no transition at all, what is consistent with the absence of the fully-localized state. The existence of this $V_{min}$ for full localization explains the distinct behavior of $C=C_C$ for small $V$, observed in Fig.\ref{fig1}-a.

\begin{figure}
      \centering
          \includegraphics[scale=0.29]{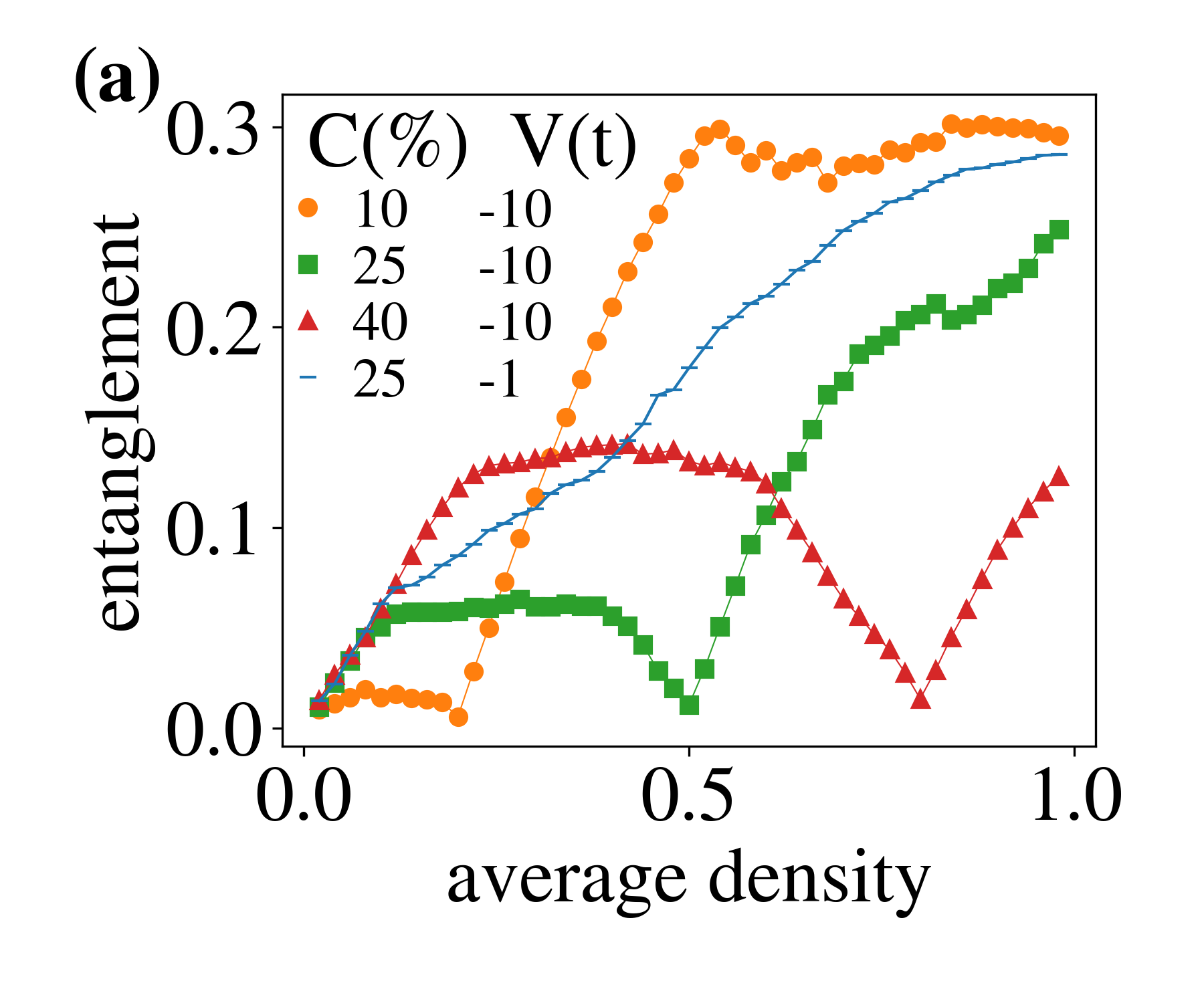} \hspace{-0.4cm}\includegraphics[scale=0.29]{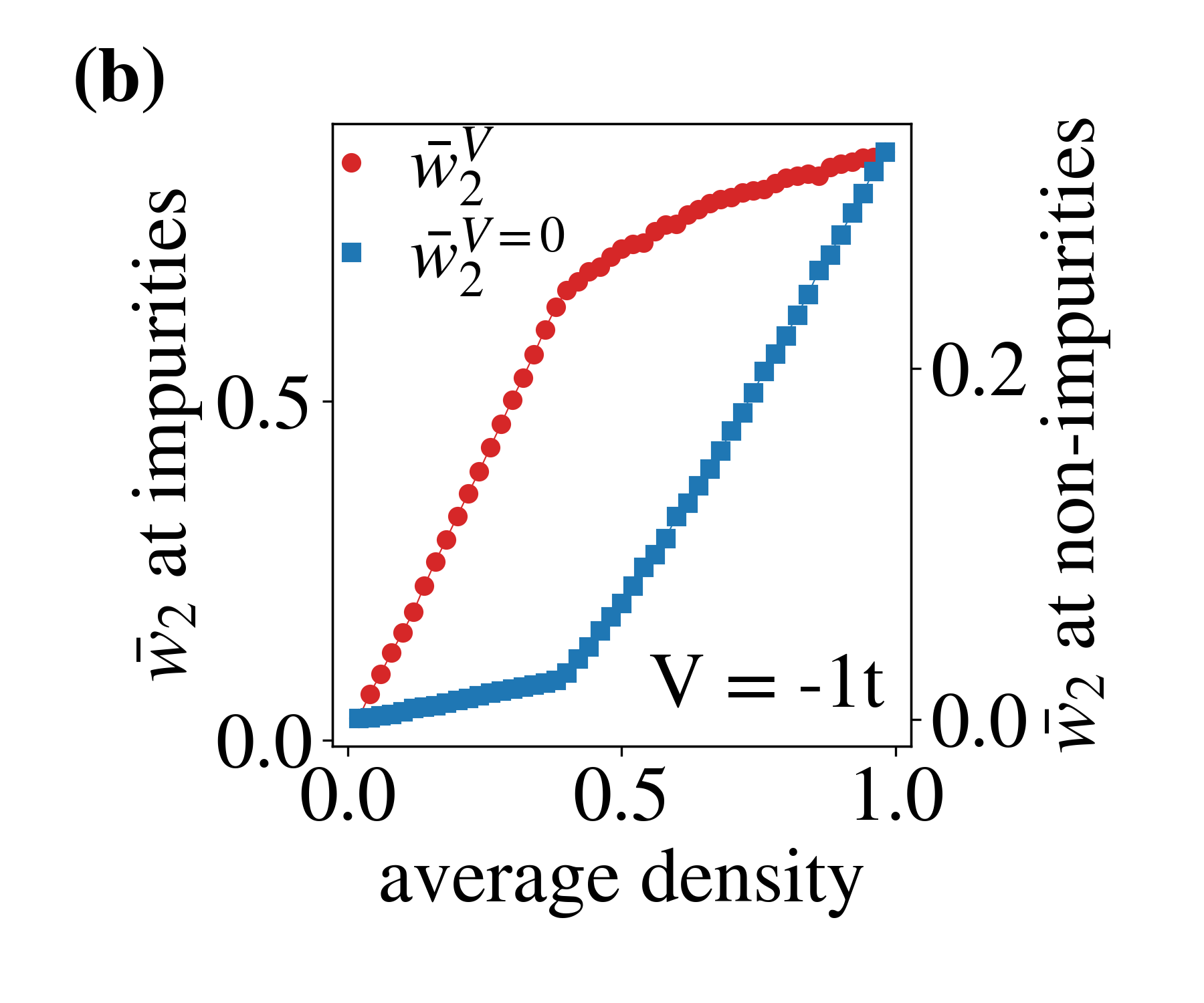}\\
       \vspace{-0.2cm}
    \includegraphics[scale=0.29]{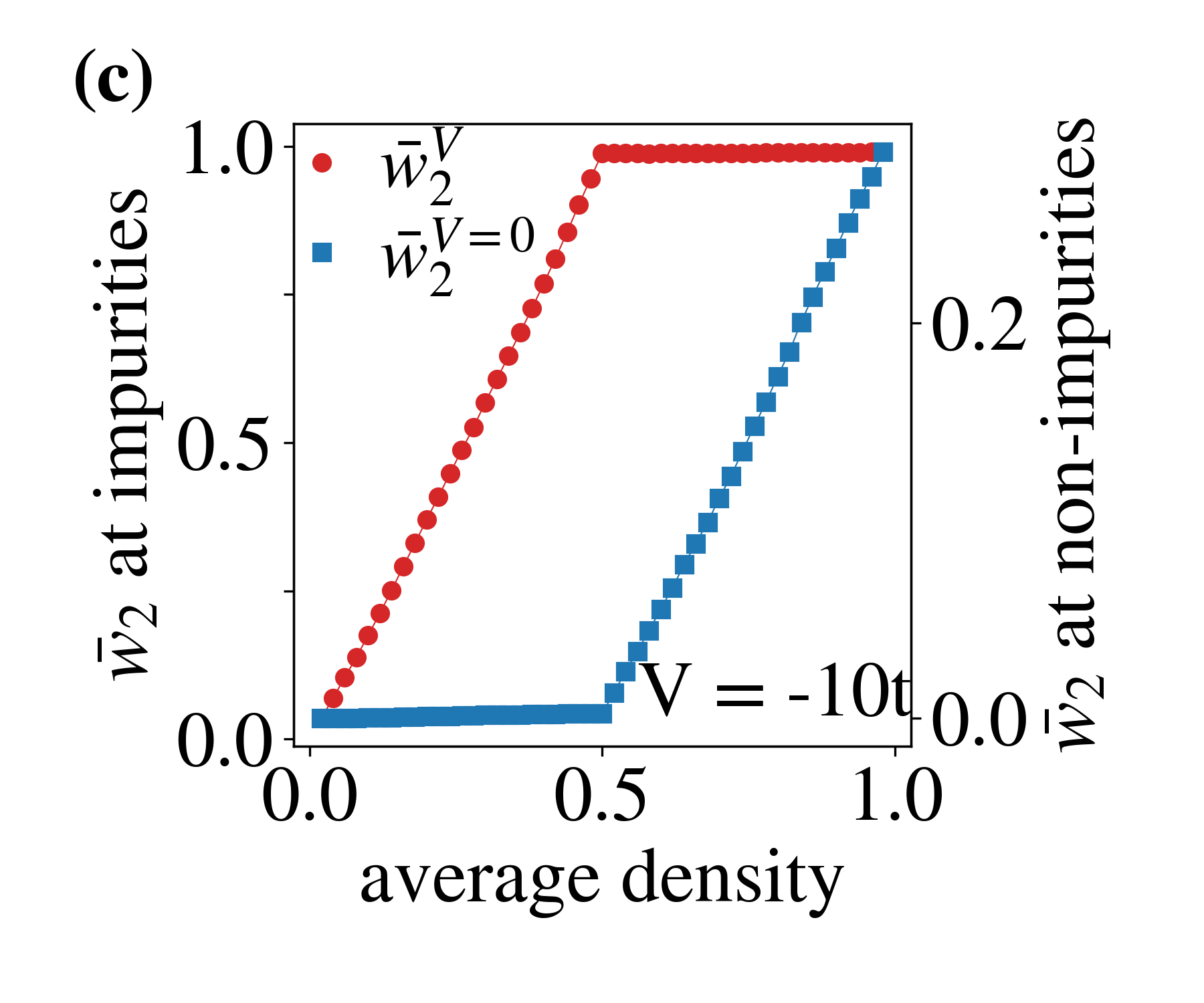}\hspace{-0.4cm} \includegraphics[scale=0.29]{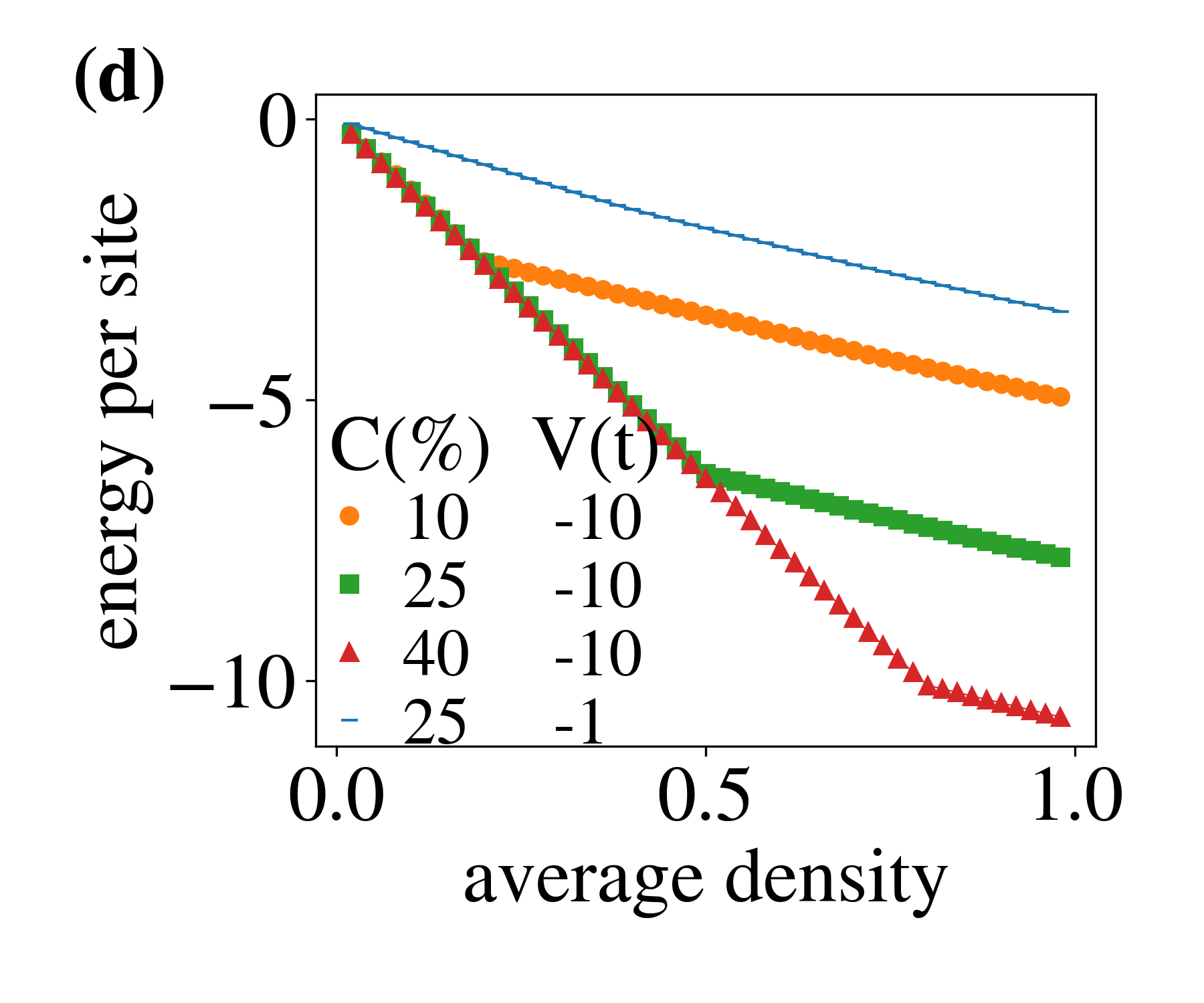}\\
    \vspace{-0.4cm}
      \caption{SIT driven by the particle density $n$: entanglement $\mathcal{L}$ (a), average double-occupation probabilities at impurity ($\bar w_2^{V}$) and non-impurity ($\bar w_2^{V=0}$) sites for $V=-t$ (b) and $V=-10t$ (c), and per-site ground-state energy (d).}\label{fig3}
   \end{figure}

{\it SIT driven by particle density.}$-$Finally, we track the transition across the average particle density for weak and strong disorder 
intensities, as shows Figure \ref{fig3}-a. For sufficiently strong $V$, all the features observed before at $C_C=100(n/2)\%$ are now found at the critical density $n_C= 2C/100$: entanglement has a sudden non-monotonicity at $n=n_C$, the system is fully localized at $n_C$ with $\mathcal{L}\approx0$, and ordinarily 
localized with  constant $\mathcal{L}>0$ for $n<n_C$. We find however that, from the fully-localized state (at $n=n_C$) to the ordinary localization (for $n<n_C$), entanglement does not increase much and has a plateau. This reflects the fact that, for small densities, disorder hampers the connections among 
the localized dimers  \cite{chinesRef12,prlRef24} due to the larger average distance among them. Consistently, the plateau is broader and with higher $\mathcal{L}$ for larger concentrations, since the average distance among dimers decreases. 

For weak disorder, $V=-t$, none of the features is observed, entanglement simply decreases monotonically with the density. This confirms the existence 
of a minimum disorder intensity $V_{min}$ for the full localization also driven by $n$. Both, the double occupancy and the energy, shown in 
Figures \ref{fig3}-b, \ref{fig3}-c and \ref{fig3}-d, corroborate the fact that the system is successfully driven by $n$ to the fully-localized state for 
$V=-10t$, while there is a frustration of the full localization for weak disorder ($V=-t$). The discontinuity at the first energy derivative \cite{sup} reveals that the SIT driven by $n$ is also a first-order quantum phase transition when leading to the fully-localized state. While the ordinary localization emerging from the SIT driven by $n$ is a smoother transition. 

The double occupancy is also useful to distinguish the nature of the SIT itself: the probabilities change smoothly {\it i)} when the SIT is driven by $V$ (Figs.\ref{fig1}-b and \ref{fig1}-c) and {\it ii)} for weak disorder, $|V|<V_{min}$, when the SIT is driven by $n$ or $C$ (Figs.\ref{fig2}-b and Fig.\ref{fig3}-b). In contrast the probabilities change abruptly for strong disorder strength with the SIT driven either by $C$ (Fig.\ref{fig2}-c) or $n$ (Fig.\ref{fig3}-c). 

In conclusion, our results may explain for example why previous works did not find an unambiguous signature of the SIT when using entanglement as witness: they have focused on the SIT driven by the disorder 
strength $V$, which has been proved here to be a smoother transition, even when leading the system to the fully-localized state. Also current controversial debates concerning the nature of the SIT and the existence or not of a critical disorder strength may be elucidated in the light of our results: the transition classification and the existence of the critical $V$ depend whether the system is led to ordinary or full localization. The impact of density, interaction and harmonic confinement $-$ essential for simulating cold atoms 
experiments $-$ on the SIT is subject of Ref. \cite{pra}. Non-local functionals for the linear entropy may be 
implemented to investigate the SIT via DFT calculations in more general potential landscapes. 

We thank Irene D'Amico for fruitful discussions. VVF was supported by FAPESP (Grant: 2013/15982-3) and GAC by the Coordena\c{c}\~{a}o de Aperfei\c{c}oamento de Pessoal de Nivel Superior - Brasil (CAPES) - Finance Code 001.


\begin{thebibliography}{99}


\bibitem{anderson} P. W. Anderson, Phys. Rev. {\bf109}, 1492 (1958).

\bibitem{supLoc} Bulaevskii, L. N. and Sadovskii, M. V. Localization and superconductivity. Pis'ma Eksp. Teor. Fiz. Zh. {\bf39}, 524 
	(1984); JETP Lett. {\bf39}, 640 (1984).

\bibitem{chinesRef12} S. Liu, X.-F. Zhou, G.-C. Guo and Y.-S. Zhang, Sci. Rep. {\bf6}, 22623 (2016); X. Cai, L.-J. Lang, S. Chen, and Y. Wang, Phys. Rev. Lett. {\bf110}, 176403 (2013); Y. Dubi, Y. Meir, and Y. Avishai, Nature {\bf449}, 876 (2007).


\bibitem{benjaminRef1} B. Sac\'ep\'e, T. Dubouchet, C. Chapelier, M. Sanquer, M. Ovadia, D. Shahar, M. Feigel'man and L. loffe, Nature 
	Physics {\bf7}, 239, (2011).
		
\bibitem{crossoverBCSBEC} A. Khan, A. Basu, and B. J. Tanatar, Supercond. Nov. Magn. {\bf26}, 1891 (2013).

\bibitem{mbl2019} N. Y. Halpern, C. D. White, S. Gopalakrishnan, and G. Refael,
Phys. Rev. B {\bf99}, 024203 (2019); R. Berkovits, Phys. Rev. B {\bf 97}, 115408 (2018); A. Acin et al., New J. Phys. {\bf 20}, 080201 (2018); J.-Y. Choi, S. Hild, J. Zeiher, P. Schauss, A. Rubio-Abadal, T. Yefsah, V. Khemani, D. A. Huse, I. Bloch, and C. Gross, Science {\bf352}, 1547 (2016); R. Stano and P. Jacquod, Nat. Photonics {\bf7}, 66 (2013).

\bibitem{benjamin} D. M. Basko, I. L. Aleiner and B. L. Altshuler, Ann. Phys. {\bf321}, 1126 (2006); V. Oganesyan and D. A. Huse, Phys. Rev. B {\bf75}, 155111 (2007); I. V. Gornyi, A. D. Mirlin, and D. G. Polyakov, Phys. Rev. Lett. {\bf95}, 206603 (2005).


\bibitem{benjaminR} A. Bezryadin, C. N. Lau, and M. Tinkham, Nature {\bf404}, 971 (2000); L. Sanchez-Palencia, and M. Lewenstein, M. Nature Phys. {\bf6}, 87 (2010); J. E. Lye, L. Fallani, M. Modugno, D. S. Wiersma, C. Fort, and M. Inguscio, Phys. Rev. Lett. {\bf95}, 070401 (2005); B. Sundar, B. Gadway, and K. R. A. Hazzard, Sci. Rep. {\bf8}, 3422 (2018).

\bibitem{ref2-32}  F. Jendrzejewski, A. Bernard, K. Muller, P. Cheinet, V. Josse, M. Piraud, L. Pezze, L. Sanchez-Palencia, A. Aspect and 
	P. Bouyer, Nature Phys {\bf8}, 398 (2012); M. Pasienski, D. McKay, M. White, and B. DeMarco, Nature Phys, {\bf6}, 677 (2010); J. Billy, V. Josse, Z. Zuo, A. Bernard, B. Hambrecht, P. Lugan, D. Clement, L. Sanchez-Palencia, P. Bouyer, and 
	A. Aspect, Nature {\bf453}, 891 (2008); G. Roati, C. DErrico, L. Fallani, M. Fattori, C. Fort, M. Zaccanti, G. Modugno, M. Modugno, M. Inguscio, Nature 
	{\bf453}, 895 (2008); L. Fallani, J. E. Lye, V. Guarrera, C. Fort, and M. Inguscio, Phys. Rev. Lett. {\bf98}, 130404 (2007).
	
	\bibitem{carterDisorder} J. M. Carter, and A. MacKinnon, Phys. Rev. B {\bf72}, 024209 (2005).

\bibitem{qpt}R. Horodecki, P. Horodecki, M. Horodecki, and K. Horodecki, Rev. Mod. Phys. 81, 865 (2009); K. Modi, A. Brodutch, H. Cable, T. Paterek, and V. Vedral, ibid. 84, 1655 (2012); V. V. Fran\c{c}a, Phys. A {\bf475}, 82 (2017); J. P. Coe, V. V. Fran\c{c}a, and I. D'Amico, EPL {\bf93}, 10001 (2011).

\bibitem{berkovitsRef13} R. Berkovits, Phys. Rev. Lett. {\bf108}, 176803 (2012); A. Zhao, R.-L. Chu, and S.-Q. Shen, Phys. Rev. B {\bf87}, 205140 (2013); V. Vettchinkina, A. Kartsev, D. Karlsson, and C. Verdozzi, Phys. Rev. B {\bf87}, 115117 (2013); H. Wang, and S. Kais, Int. J. Quantum Inform. {\bf4}, 827 (2006).

\bibitem{islamRef18} R. Islam, R. Ma, P. M. Preiss, M. E. Tai, A. Lukin, M. Rispoli, and M. Greiner, Nature {\bf528}, 77 (2015); N. Laflorencie, Phys. Rep. {\bf646}, 1 (2016); A. M. Goldsborough, and R. A. R\"omer, EPL {\bf111}, 26004 (2015); X. Deng, R. Citro, E. Orignac, A. Minguzzi, and L. Santos, New J. of Phys. {\bf15}, 045023 (2013);  I. Fr\'erot, and T. Roscilde, Phys. Rev. Lett. {\bf116}, 190401 (2016).

\bibitem{albusRef37} A. Albus, F. Illuminati, and J. Eisert, Phys. Rev. A {\bf68}, 023606 (2003).

\bibitem{royRef14} N. Roy, and A. Sharma, Phys. Rev. B {\bf97}, 125116 (2018); B.-T. Ye, Z.-Y. Han, L.-Z. Mu, and H. Fan, Sci. Rep. {\bf7}, 16668 (2017); I. Mondragon-Shem, M. Khan, and T. L. Hughes, Phys. Rev. Lett. {\bf110}, 046806 (2013).

\bibitem{buchholdRef16} A. Kapitulnik, S. A. Kivelson, and B. Spivak, Rev. of Mod. Phys. {\bf91}, 011002 (2019); M. Buchhold, S. Diehl, and A. Altland, Phys. Rev. Lett. {\bf121}, 215301 (2018); J. H. Pixley, D. A. Huse, and S. Das Sarma, Phys. Rev. B {\bf94}, 121107 (2016); K. Ziegler, and A. Sinner, Phys. Rev. Lett. {\bf121}, 166401 (2018); M. Tezuka, Phys. Rev. A {\bf82}, 043613 (2010).
	
	\bibitem{bouRef4} K. Bouadim, Y. L. Loh, M. Randeria, and N. Trivedi, Nature Phys. {\bf7}, 884 (2011).



\bibitem{dft} W. Kohn, Rev. Mod. Phys. {\bf71}, 1253 (1999); W. Kohn, and L. J. Sham, Phys. Rev. {\bf140}, 1133 (1965); K. Capelle, and V. L. Campo Jr., Phys. Rep. {\bf528}, 91 (2013). J. P. Coe, I. D'Amico, V. V. Fran\c{c}a, EPL {\bf110}, 63001 (2015). V. V. Fran\c{c}a, D. Vieira, and K. Capelle, New J. of Phys. {\bf14}, 073021 (2012).


\bibitem{francaAmico2011} V. V. Fran\c{c}a, and I. D'Amico, Phys. Rev. A {\bf83}, 042311 (2011).

\bibitem{sup1} See Supplemental Material at [URL will be inserted by publisher] for the entanglement convergence as a function of the number of samples.

\bibitem{prl2008} V. V. Fran\c{c}a and K. Capelle, Phys. Rev. Lett. {\bf100}, 070403 (2008).


\bibitem{praRef25} Y. Cao, G. Xianlong, X.-J. Liu, and H. Hu, Phys. Rev. A {\bf93}, 043621 (2016); R. T. Scalettar, N. Trivedi, and C. Huscroft, Phys. Rev. B {\bf59}, 4364 (1999).

\bibitem{pra} G. A. Canella and V. V. Fran\c{c}a, {\it Joint Publication} submitted to Phys. Rev. A (2019).

\bibitem{sup} See Supplemental Material at [URL will be inserted by publisher] for the energy derivatives.
\bibitem{sachdev} S. Sachdev, Quantum phase transitions (Cambridge University Press, Cambridge, 2011).

\bibitem{prlRef24} S. Ospelkaus, C. Ospelkaus, O. Wille, M. Succo, P. Ernst, K. Sengstock, and K. Bongs, Phys. Rev. Lett. {\bf96}, 180403 
	(2006).



\end{thebibliography}
\end{document}